\documentclass[a4paper, 11pt, onecolumn, sort&compress]{elsarticle}

\usepackage{graphicx}
\usepackage{amssymb}
\usepackage{bm}
\usepackage{booktabs}
\usepackage{amsmath}
\usepackage{amsthm} 
\usepackage{nccmath}
\usepackage{array}
\usepackage{color}
\usepackage{varioref}
\usepackage{enumerate} 
\usepackage{geometry}
\usepackage{algorithm}
\usepackage{algorithmic}
\usepackage{fullwidth}
\usepackage{xcolor}
\usepackage{tikz}
\usepackage{comment}
\usepackage{geometry}
\usepackage{multirow}
\usepackage{diagbox}
\usepackage{appendix}
\usepackage{lineno}

\newcommand{\tikzmark}[1]{\tikz[remember picture] \node[coordinate, anchor = west] (#1) {#1};}
\numberwithin{equation}{section}

\makeatletter
\let\start@align@nopar\start@align
\let\start@gather@nopar\start@gather
\let\start@multline@nopar\start@multline
\long\def\start@align{\par\start@align@nopar}
\long\def\start@gather{\par\start@gather@nopar}
\long\def\start@multline{\par\start@multline@nopar}
\makeatother

\newtheorem{example}{Example}

\newcommand{\bs}[1]{\ensuremath{\boldsymbol{#1}}}
\newcommand{\mc}[1]{\ensuremath{\mathcal{#1}}}
\newcommand{\abs}[1]{\ensuremath{\left| \ #1 \ \right|}}

\newcommand{\R}{\ensuremath{\mathbb{R}}}

\newcommand{\Np}{\ensuremath{N_{\mathcal{P}}}}
\newcommand{\Ntop}{\ensuremath{N_{\text{top}}}}

\newcommand{\p}[1]{\ensuremath{\mathbb{P}\left(#1\right)}}

\newcommand{\E}[1]{\ensuremath{\mathbb{E}\left[#1\right]}}

\newcommand{\Var}[1]{\ensuremath{\mathbb{V}\text{ar}\left(#1\right)}}

\DeclareMathOperator*{\argmax}{arg\,max}
\DeclareMathOperator*{\argmin}{arg\,min}

\begin{document}
\begin{frontmatter}
  \title{DNA mixture deconvolution using an evolutionary algorithm with multiple populations, hill-climbing, and guided mutation}
  
\author[aau]{S\o ren B. Vilsen\corref{math}}
\cortext[math]{Skjernvej 4A; DK-9220 Aalborg East; Denmark}
\ead{svilsen@math.aau.dk}  
\author[aau]{Torben Tvedebrink}
\author[aau]{Poul Svante Eriksen}

\address[aau]{Department of Mathematical Sciences, Aalborg University, Denmark}
  \begin{abstract}
DNA samples crime cases analysed in forensic genetics, frequently contain DNA from multiple contributors. These occur as convolutions of the DNA profiles of the individual contributors to the DNA sample. Thus, in cases where one or more of the contributors were unknown, an objective of interest would be the separation, often called deconvolution, of these unknown profiles. In order to obtain deconvolutions of the unknown DNA profiles, we introduced a multiple population evolutionary algorithm (MEA). We allowed the mutation operator of the MEA to utilise that the fitness is based on a probabilistic model and guide it by using the deviations between the observed and the expected value for every element of the encoded individual. This guided mutation operator (GM) was designed such that the larger the deviation the higher probability of mutation. Furthermore, the GM was inhomogeneous in time, decreasing to a specified lower bound as the number of iterations increased. This ensured that the operator would not fixate on the elements with large deviance residual, which could not be improved.

We analysed 102 two-person DNA mixture samples in varying mixture proportions. The samples were quantified using two different DNA prep. kits: (1) Illumina ForenSeq Panel B (30 samples), and (2) Applied Biosystems Precision ID Globalfiler NGS STR panel (72 samples). The DNA mixtures were deconvoluted by the MEA and compared to the true DNA profiles of the sample. We analysed three scenarios where we assumed: (1) the DNA profile of the major contributor was unknown, (2) DNA profile of the minor was unknown, and (3) both DNA profiles were unknown. Furthermore, we conducted a series of sensitivity experiments on the ForenSeq panel by varying the sub-population size, comparing a completely random homogeneous mutation operator to the guided operator with varying mutation decay rates, and allowing for hill-climbing of the parent population. The sensitivity experiments showed that if the mutation was guided, then the mutation decay rate and the number of hill-climbing iterations had little to no effect on the outcome. While if the mutation was random, then hill-climbing was a valuable addition to MAE, in order to achieve an equivalent precision. Furthermore, the sensitivity study showed that vary the number of sub-populations did not change the precision of the results, but yielded for faster convergence as the sub-populations were easily parallelised.

The deconvolution of the Illumina ForenSeq Panel B samples showed that if the profile of the minor contributor was known the MAE found the profile major with a high accuracy, while if the profile of the major contributor was known, when finding the minor was more difficult, but still had a precision above 85\% in 3:1 mixtures. If both profiles were unknown the mixture had to be skewed in order to achieve similar results. The results of the deconvolution of Applied Biosystems Precision ID Globalfiler NGS STR samples were similar in general and further showed that as the total amount of input DNA decreased identification of the minor profile became only slightly more difficult.

\end{abstract}
  
\end{frontmatter}

\section{Introduction}
A central question when DNA evidence, $\mc E$, is presented in court is the determination of the posterior odds of the evidence under the competing hypotheses of the prosecution, $\mc H_p$, and defence, $\mc H_d$. Given DNA evidence, a DNA profile can be created, even from small samples of a few hundred pico-grams. A DNA profile is created by examining locations of the DNA, called markers. The markers are chosen such that they have a large variation in the population, but have a low mutation rate. At each (autosomal) marker a person may take two values, called alleles, and the collection of these alleles constitutes the persons DNA profile. If the sample is in large quantity and contains DNA from a single contributor, then identification is simple. However, when the DNA is found at a crime scene, the sample can be contaminated, be in extremely low quantity, contain DNA from multiple contributors, or any combination thereof \cite{Butler2009, Butler2012}. If the sample contains DNA from multiple contributors, called a DNA mixture, then accurately representing the weight-of-evidence against a suspects guilt versus its innocence (the posterior odds) is very difficult, because the number of contributors, their relative contribution to the mixture, their DNA profiles, etc., are all unknowns. 

In general the posterior odds of the two hypotheses can, by applying Bayes' theorem and denoting a probability by $\p{\cdot}$, be written as:
\begin{align*}
\frac{\p{\mc H_d|\mc E}}{\p{\mc H_p|\mc E}} = \frac{\p{\mc E|\mc H_d}}{\p{\mc E|\mc H_p}}\frac{\p{\mc H_d}}{\p{\mc H_p}}.
\end{align*}

The prior odds of the two hypotheses should be supplied by the court, as it should represent the odds of the two hypotheses based on the evidence introduced to the court prior to the introduction of the DNA evidence. This leaves the likelihood ratio:
\begin{align*}
\text{LR}(\mc H_d, \mc H_p) = \frac{\p{\mc E|\mc H_d}}{\p{\mc E|\mc H_p}},
\end{align*}
for which, we write $\text{LR}$ in the remainder of the manuscript.

The prevailing method of analysing DNA evidence is by determining the length of short tandem repeat (STR) regions using capillary electrophoresis (CE). In recent years, massively parallel sequences (MPS) has started to be introduced in forensic genetics casework. MPS offers not just the length, but the entire base composition of the STR regions. That is, DNA samples analysed with MPS will be of higher resolution than those analysed with CE \cite{paper4:Borsting2015}. 

When a DNA sample is analysed it will yield some quantitative information $\bs y$ (called the coverage in the MPS setting) about some combined genetic information $\bs g_c$, structurally specified by a hypothesis $\mc H_i$. Therefore, we can factorise $\p{\mc E|\mc H_i}$ as: 
\begin{align*}
\p{\mc E|\mc H_i} = \p{\bs y, \bs g_c|\mc H_i} = \p{\bs y|\bs g_c}\p{\bs g_c|\mc H_i}.
\end{align*}

That is, the probability of the evidence, given a hypothesis, has two parts: (1) The probability of the quantitative information given the genetic information, and (2) the probability of the genetic information under the hypothesis, which is usually assumed to follow a Dirichlet-Multinomial distribution, as described in Balding and Nichols (1994) \cite{paper4:Balding1994}. In this paper, the quantitative information will be given as the two component model described in Section \ref{sec:app:mixturemodel}.

It is suppressed that the probability of the quantitative information given the genetic information, $\p{\bs y|\bs g_c}$, will depend on unknown parameters, $\bs \theta$, describing the uncertainty of the measuring process, the signal intensities and the mixture proportions.

If a hypothesis, $\mc H_i$, states that a contributor is unknown it is necessary to sum over the set of possible unknown contributors, $\mc U$, to obtain the probability of the evidence. That is, if we assume (1) the sample contains DNA from two contributors, (2) one DNA profile is known, $\bs g_k$ (this could e.g.\ be the victim), and (3) $\mc H_i$ states that the DNA is a mixture of the known profile and a random unknown profile from the population. The evidence under the hypothesis is $\mc E = (\bs y, \bs g_k)$ and probability of the evidence reduces to:
\begin{align}
\p{\bs y, \bs g_k|\mc H_i} \approx \sum_{\bs g\;\in\;\mc U} \p{\bs y | \bs g_k, \bs g, \hat{\bs \theta}_{\mc U}}\p{\bs g|\bs g_k}, \label{eq:intro:poe}
\end{align} 
where $\hat{\bs \theta}_{\mc U}$ is the $\bs \theta$ maximising the sum of the probabilities: $\sum_{\bs g\;\in\;\mc U} \p{\bs y | \bs g_k, \bs g, \bs \theta}$. The notation $\hat{\bs \theta}_{\mc U}$ should be interpreted as: the parameters were dependent on the entirety of the evidence, i.e.\ including every possible unknown genotype. Note that this formulation of the probability of the evidence can be extended to an arbitrary number of known and unknown DNA profiles. 

The sum over the set of unknown contributors, $\mc U$, may be intractable because of the size of $\mc U$. In the CE setting, this problem is largely solved by using a Bayesian network \cite{paper4:cowell_etal_2015, paper4:oyvind_bleka_2016}, by sampling from the posterior distribution using Markov chains \cite{paper4:taylor_etal_2013}, or by simply limiting the number of unknown contributors to cases where the sum is tractable \cite{paper4:tvedebrink_etal_2012, paper4:steele_2016}. However, as MPS is still relatively immature in the forensic genetics setting, the methods used for analysing CE DNA mixtures cannot be applied directly to MPS data, at least not without modification \cite{paper4:Vilsen2018b}. 

The set $\mc U$ is discrete and does not have any natural ordering, making an Evolutionary Algorithm (EA) a perfect tool \cite{paper4:Blum2003, paper4:Tu2004, paper4:Das2011, paper4:Gogna2013, Campos1999, Larranaga2013} for finding the combination of unknown genotypes maximising: 
\[\p{\bs y | \bs g_k, \bs g, \hat{\bs \theta}_{\bs g}}\p{\bs g|\bs g_k},\] 
w.r.t. $\bs g$ (the subscript in $\hat{\bs \theta}_{\bs g}$ indicates that the estimated parameters only depended on the unknown profile $\bs g$ and not the entire set $\mc U$). Identifying the optimal unknown profile combination is of interest as it could be used as a starting point when searching a DNA database for a potential suspect, in cases where the investigators had no other leads. The main advantage to using EA's compared to the sampling methods employed in e.g.\ \cite{paper4:taylor_etal_2013}, is the crossover operator, as it allows for larger changes to the proposed individual and, thus, it can more effectively search the set of unknown genotype combinations.

In most EA's \cite{Campos1999, Larranaga2013}, the mutation operator is completely random, using a flat mutation rate (which may be inhomogeneous in time) for every element of an individual. We will refer to this as the random mutation operator (RM). The main advantage of the RM is that it searches the state space thoroughly. However, if the state space is large relative to the set of optimal solutions, then most of the steps taken by the RM will often lead to only small increases to the solution currently proposed by the algorithm.

We propose to guide the mutation by augmenting the probability of an element (of an individual) mutating based on the deviation seen between the observed data and what is expected given the decoded individual. We will refer to this mutation operator as the guided mutation operator (GM). The GM was designed to have a larger chance of making mutation which will increase the fitness of the individual, at the cost of not searching the state space as thoroughly as the RM and being more quickly fixated at a local optimum. In order to compensate for these two disadvantages, we added multiple randomly initialised sub-populations, and allowed for migration between these sub-population \cite{paper4:Muhlenbein1991, paper4:Muhlenbein1991b}. 

The manuscript is organised as follows: Section \ref{sec:app:mixturemodel} contains a short introduction to a two component DNA mixture coverage model, followed by Section \ref{sec:mea}, where we introduce the general structure of the MEA algorithm and its operators. Section \ref{sec:experiments} contains short description of the data to be analysed, a sensitivity study, and an examination of performance of the MEA implementation. Lastly, concluding remarks are given in Section \ref{sec:conc}. 

Lastly, the MEA described below was implemented in \texttt{R} and \texttt{C++} through the \texttt{Rcpp}-packages using the Eigen, Boost, and NLopt libraries \cite{paper4:R, paper4:Cpp, paper4:Rcpp, paper4:CppEigen, paper4:RcppEigen, paper4:Boost, paper4:BH, paper4:NLopt, paper4:NLoptR} in the \texttt{R}-package MPSMixtures \cite{paper4:MPSMixtures}.

\section{The two component DNA mixture coverage model}\label{sec:app:mixturemodel}
The probabilistic coverage model has two two parts: (1) an allele coverage model, which will account for observed alleles, and (2) a noise coverage model accounting for the remaining observations.

The allele coverage model was introduced in \cite{paper4:Vilsen2018a, paper4:Vilsen2018b} and is an adaptation of the capillary electrophoresis gamma peak height models \cite{paper4:Cowell2011, paper4:cowell_etal_2015}. The coverage of a string is a synonym for the number of times the string was observed in the sample. Therefore, it is natural to model the coverage of an allele as overdispersed count data. Furthermore, preliminary analyses showed that the variance of high coverage alleles were too large when using the PG2 model. Therefore, we will assume that the coverage of marker $m$ allele $a$, denoted $y_{ma}$, follows a PG1 distribution, see e.g.\ Appendix \ref{app:poissongamma}: 
\begin{align}
y_{ma} \sim \text{PG1}\left(\mu_{ma}, \gamma\right),
\end{align}
where $\gamma$ is the overdispersion and $\mu_{ma}$ is the expected coverage defined as:
\begin{align}
\mu_{ma} = \nu \beta_m \sum_{c = 1}^C \left[g_{mac} + s^{(k)}_{mac}\right]\varphi_c,
\end{align}
where $\nu$ can be interpreted as the average heterozygote coverage (when there is only a single contributor to the mixture), $\beta_m$ is a marker dependent scaling parameter, $C$ is the total number of contributors, $\varphi_c$ is the relative contribution of contributor $c$, with $\sum_{c = 1}^C \varphi_c = 1$, $g_{mac}\in \{0, 1, 2\}$ is the count of allele $a$ contributor $c$ has on marker $m$, and $s^{(k)}$ is defined as:
\begin{align*}
s^{(k)}_{mac} = \sum_{A \in \mc P(a)} \xi_{mA} \left(g_{mAc} + s^{(k-1)}_{mAc}\right).
\end{align*}
where $\mc P(a)$ is the set of potential parents of allele $a$, $\xi_{mA}$ is the stutter ratio of parent $A$, and $s^{(0)} = 0$. For most practical applications $k \leq 3$ (and in this paper $k = 2$), as any contributions from the fourth level would be minuscule. To see how small, assume that a sequence at the fourth level has a coverage of 1,000 and that we have a stutter ratio of 0.15 in all four levels (which is a high stutter ratio), then coverage received from this fourth level sequence would be close to a half, as opposed to 1,000.

The marker imbalances, $\bs \beta$, are estimated using the convex combination of a maximum likelihood estimate and method-of-moments estimate: 
\begin{align}\label{eq:papermix:materials:updated:convex}
\tilde{\bs\beta}_{\lambda} = \lambda \hat{\bs \beta} + (1 - \lambda) \bs\beta^{\text{MoM}},\;\;\text{where}\;\;\lambda\in[0;1],
\end{align}
where $\hat{\bs \beta}$ are MLE,estimated using a database of samples \cite{paper4:Vilsen2018b}, and $\bs{\beta}^{\text{MoM}}$ are the following method-of-moments (MoM) estimate:
\begin{align*}
\beta_m^{\text{MoM}} = \dfrac{\sum\limits_a y_{ma}}{\frac{1}{_M}\sum\limits_{m, a} y_{ma}}.
\end{align*}
Note: that the MoM estimate only depends on the sample itself, while the MLE needs a database of samples sequenced with the same technology as the sample to be analysed \cite{paper4:Vilsen2018b}.

Lastly, $\lambda$ represents the weight/belief we assign the workflow database. The closer $\lambda$ gets to 1, the more weight we put on the workflow database based estimates, $\hat{\bs \beta}$. Whereas the closer it gets to 0, the more emphasis we put on the moment based estimates, $\bs \beta^{\text{MoM}}$. Furthermore, in this paper, the parameters $\hat{\bs\beta}$, the parameters $\bs\xi$, and the set of potential parents were assumed known. 

Any observation not identified as an allele or a stutter of an allele is classified as noise and denoted using set complement notation: $\bs y^c$ and $\bs g^c$. The coverage of the noise is assumed to follow a one-inflated, zero-truncated Poisson-gamma model with mean $\mu$ and overdispersion $\rho$  \cite{Vilsen2015, paper4:Vilsen2017}. The noise distribution has to be truncated at zero, as a coverage of zero implies that the string has not been observed, which implies that it cannot be noise. The one-inflation is necessary to allow for shoulders, back stutters, or higher order stutters, in the noise distribution, without increasing the variance to such a degree that it would make noise coverage model meaningless. Meaningless in the sense that it would be able to fit literally anything.

The complete log-likelihood of the two component model, can be written as:
\begin{align}
\begin{split}
&\ell(\bs\theta | \bs \beta, \bs\xi, \bs y, \bs g) = \\
&\quad \sum_m \ell(\nu, \eta, \bs \varphi | \beta_m, \bs\xi_m, \bs y_m, \bs g_m) + \ell(\mu, \rho | \bs y^c_m, \bs g^c_m),
\end{split} \label{eq:app:mixtureloglikelihood}
\end{align}
as the two components are assumed independent given the genotypic information.

\section{The multiple population evolutionary algorithm}\label{sec:mea}
We chose an MEA for the following two reasons: (1) if we initialise these populations randomly, then they will more thoroughly explore the fitness landscape, when compared to running a single populations algorithm, and (2) it allows us to utilise a smaller total population size, thereby, decreasing runtime. This implementation is a variation of the parallel evolutionary algorithm presented by M\"{u}hlenbein et al. \cite{paper4:Muhlenbein1991, paper4:Muhlenbein1991b}.

An outline of the implemented MEA can be seen in Algorithm~\ref{alg:mea}. The MEA works by segregating the total population, $\mc P$, into $\Np$ smaller sub-populations, $\bs P_n$, each containing $N_I$ individuals. The $i$'th individual of sub-population $n$ will be referred to $\bs p_{ni}$, dropping subscripts if they are unnecessary or clear from context.  

A single iterations of the MEA, consists of two phases: 
\begin{itemize}
\item[(1)] the migration phase (Algorithm~\ref{alg:mea}: Line 4).
\item[(2)] the evolution phase (Algorithm~\ref{alg:mea}: Lines 6-16).
\end{itemize}

During the migration phase, the $\Np$ sub-populations exchange information according to a predefined pattern. The information exchanged and the migration pattern is described in Subsection \ref{sec:migration} below. During the second phase, an EA is run on each sub-population independently, by in turn hill-climbing each individual in the population creating a parent, find a partner to the parent, use crossover to create a child of the individuals, mutate the child, and determine whether the child should replace its parent. A more detailed description of the selection, representation, and operators of the independent EAs can be found in Subsections \ref{sec:selection} and \ref{sec:operators} below. Note that in the entirety of this manuscript the word individual will always be used in the EA context, i.e.\ an individual can describe the genetic profile of multiple contributors (persons), whereas as person will always be referred to as a contributor.

Lastly, we updated the individual of largest fitness, $\hat{\bs g}$. The implemented MEA is said to have converged when the difference between the individuals of largest fitness of each sub-population is smaller than some $\varepsilon$ for more than $N_{\varepsilon}$ iterations. 

Note that the functions \texttt{crossover}, \texttt{mutate}, and \texttt{hillclimb} will be introduced in Algorithms \ref{alg:crossover}, \ref{alg:mutation}, and \ref{alg:hillclimbing}, respectively.

\begin{algorithm}
\caption{\label{alg:mea}Multiple Population Evolutionary Algorithm.}
\begin{algorithmic}[1]
\renewcommand{\algorithmicrequire}{\textbf{Input:}}
\renewcommand{\algorithmicensure}{\textbf{Output:}}
\REQUIRE $\bs y$, $\bs g_k$, $\varepsilon$, $N_{\mc P}$, $N_{I}$, $N_{\varepsilon}$, $\Ntop$
\ENSURE $\hat{\bs g}$
\STATE $\mc P$: randomly initialise $N_{\mc P}$ sub-populations each with $N_{I}$ individuals.
\WHILE {$\big($not converged$\big)$} 
	\STATE $\mc P \leftarrow \texttt{migrate}(\mc P)$
	\FOR {$\big(n$ from $1$ to $N_{\mc P}\big)$}
		\STATE $\tilde{\bs P}$: empty sub-population. 
		\FOR {$\big(i$ from $1$ to $N_{I}\big)$}
			\STATE $\tilde{\bs p} \leftarrow \texttt{hillclimb}(\bs p_{ni})$
			\STATE $\bs q \leftarrow \texttt{selectpartner}(\tilde{\bs p}, \bs P_n)$
			\STATE $\bs c \leftarrow \texttt{crossover}(\tilde{\bs p}, \bs q)$.
			\STATE $\bs c \leftarrow \texttt{mutate}(\bs c)$
				\IF {$\big(F(\tilde{\bs p}) < F(\bs c)\big)$} 
					\STATE $\tilde{\bs p} \leftarrow \bs c$
				\ENDIF
			\STATE Append $\tilde{\bs p}$ to $\tilde{\bs P}$	
		\ENDFOR
		\STATE $\bs P_n \leftarrow \tilde{\bs P}$
	\ENDFOR
	\STATE Update $\hat{\bs g}$
	\STATE Update convergence criteria
\ENDWHILE 
\RETURN $\bs P_{\text{top}}$ 
\end{algorithmic}
\end{algorithm}

\subsection{Migration} \label{sec:migration}
When migration occurs the highest fitness individual of the sub-population is copied and send to its neighbours, replacing the neighbours individual of lowest fitness. The migration operator, combined with the fact that the sub-populations were randomly initialised, creates the following advantage: If a sub-population gets stuck at a local maximum, then a migration of the highest fitness individual from another sub-population can help drag it out of the local maximum (hopefully towards the global maximum, or at the very least push it towards another part of the sample space).

\begin{figure}[ht!]
\centering
\begin{tikzpicture}
\node (a) [draw, circle, minimum size = 1.25cm, inner sep = 2pt] at (0, 0) {$\bs P_1$};
\node (b) [draw, circle, minimum size = 1.25cm, inner sep = 2pt] at (2, 0) {$\bs P_2$};
\node (c) [draw, circle, minimum size = 1.25cm, inner sep = 2pt] at (4, 0) {$\bs P_3$};
\node (d) [draw, circle, minimum size = 1.25cm, inner sep = 2pt] at (5, -1.5) {$\bs P_4$};

\node (ein) [draw, circle, draw opacity = 0, minimum size = 1.25cm, inner sep = 2pt] at (4, -3) {};
\node (e) [draw, circle, draw opacity = 0, minimum size = 1.25cm, inner sep = 2pt] at (3, -3) {$\cdots$};
\node (eout) [draw, circle, draw opacity = 0, minimum size = 1.25cm, inner sep = 2pt] at (2, -3) {};

\node (f) [draw, circle, minimum size = 1.25cm, inner sep = 2pt] at (0, -3) {$\bs P_{\Np-1}$};
\node (g) [draw, circle, minimum size = 1.25cm, inner sep = 2pt] at (-1, -1.5) {$\bs P_{\Np}$};

\draw [->, >=latex] (a) to (b);
\draw [->, >=latex] (b) to (c);
\draw [->, >=latex] (c) to (d);
\draw [->, >=latex] (d) to (ein);
\draw [->, >=latex] (eout) to (f);
\draw [->, >=latex] (f) to (g);
\draw [->, >=latex] (g) to (a);

\draw [->, >=latex] (a) to (f);
\draw [->, >=latex] (b) to (g);
\draw [->, >=latex] (c) to[out = 150, in = 30] (a);
\draw [->, >=latex] (d) to (b);
\draw [->, >=latex] (ein) to (c);
\draw [->, >=latex] (g) to (eout);
\end{tikzpicture}
\caption{\label{fig:migration}The migration neighbourhood structure used in the MEA.}
\end{figure}
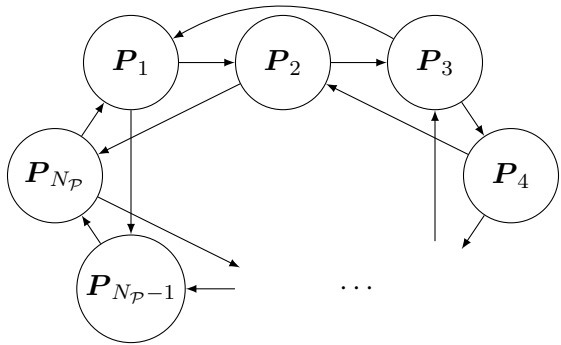

When designing the migration operator, we needed to balance the sharing of high fitness individuals, while simultaneously ensuring that the high fitness individuals did not spread too quickly, as the algorithm may then fixate at a local maximum. Therefore, we used the neighbourhood structure shown in Fig. \ref{fig:migration} (inspired by \cite{paper4:Muhlenbein1991} and \cite{paper4:Muhlenbein1991b}). 

Using this structure and assuming that an individual of higher fitness was not created during this period, the minimum number of iterations needed for an individual to spread to every sub-population is exactly: 
\begin{align}\label{eq:migration:numberiterations}
\lceil (\Np + 1) / 3\rceil,
\end{align}
which follows by counting the number of ways to get to all states in both the clockwise and anti-clockwise direction. 

\subsection{Representation and fitness} \label{sec:solution}
\subsubsection{The set of unknown genotype combinations}
Any genotype on an autosomal marker will have exactly two alleles; they can be different or equal, called hetero- and homozygous, respectively, but there will always be exactly two alleles (disregarding extremely rare events). Therefore, any genotype on a marker $m$ with $A_m$ different observed alleles, denoted $\bs g_m$, can be represented as a vector with elements $g_{mi} \in \{0, 1, 2\}^{A_m}$, where $\sum_i g_{mi} = 2$.

In order to understand the shape and size of $\mc U$ and the representation used in the EA, we start by giving a small example.

\begin{example}
Assume we have a single marker with three observed alleles and a single unknown contributor, $u$. The possible unknown genotypes of $u$ are:
\begin{align*}
\mc U_1 = &\left\{
\begin{bmatrix}
1 \\ 1 \\ 0
\end{bmatrix},\;
\begin{bmatrix}
1 \\ 0 \\ 1
\end{bmatrix},\;
\begin{bmatrix}
0 \\ 1 \\ 1
\end{bmatrix},\;\begin{bmatrix}
2 \\ 0 \\ 0
\end{bmatrix},\;
\begin{bmatrix}
0 \\ 2 \\ 0
\end{bmatrix},\;
 \begin{bmatrix}
0 \\ 0 \\ 2
\end{bmatrix}
\right\},
\end{align*}
i.e.\ a total of six possible genotypes. Note the subscript in $\mc U_1$ refers to the number of unknown contributors. 
\end{example}

In general, assuming the total number of observations on a marker $m$ is $A_m$, then the number of heterozgote and homozygote genotypes can be written as:
\begin{align*}
\binom{A_m}{2}\;\;\text{and} \;\; A_m,
\end{align*}
respectively. Thus, for $M$ independent markers, the size of $\mc U_1$ is given as follows:
\begin{align*}
\abs{\mc U_1} = \prod_{m = 1}^M \frac{A_m(A_m + 1)}{2}.
\end{align*}
Furthermore, if there are $U$ unknown contributors, then the size of $\mc U_{U}$ is 
\[\abs{\mc U_{U}} = \abs{\mc U_1}^{U}.\]

However, as the order in which the genotypes appear will not affect the fitness, the size of the set can ultimately be reduced to:
\begin{align*}
\abs{\mc U_{U}} = \binom{\abs{\mc U_1} + U - 1}{U}.
\end{align*}

\subsubsection{Representation of unknown genotypes}
We have encoded the genotypes of an unknown contributor using two elements per marker. This representation has two main advantages: (1) it simplifies the cross-over and mutation operators, and (2) it will always require less memory. 

An individual will be encoded as pointers to the non-zero elements of the genotype matrix. Because the markers are assumed to be independent, encoding will be performed on a marker-by-marker basis, taking one contributor at a time. 
\begin{example}
Continued from Example 1, but assuming we have two unknown contributors. Furthermore, assume that the genotypes of the two unknown contributors are given by the matrix: 
\begin{align}\label{eq:solution:genotypematrix}
\bs g = \begin{bmatrix}
1 & 1 \\
0 & 1 \\
1 & 0
\end{bmatrix},
\end{align}
where the first and second columns correspond to the first and second unknown contributor, respectively. Then, the encoded individual, $\bs p$, of $\bs g$ is:
\begin{align}\label{eq:solution:encoded}
\bs p = \begin{bmatrix}
1 \\ 3 \\ 1 \\ 2
\end{bmatrix},
\end{align}
where the first two elements correspond to the first contributor and the last two elements to the second, respectively, as:
\begin{align*}
\bs g = \left[\begin{array}{ll}
1\tikzmark{as} & 1\tikzmark{bs} \\ 0 & 1\tikzmark{ds} \\ 1\tikzmark{cs} & 0 \end{array} \right] \quad
\rightarrow \quad
\bs p = \left[\begin{array}{l}
	\tikzmark{ae}1 \\ \tikzmark{ce}3 \\ \tikzmark{be}1 \\ \tikzmark{de}2 
\end{array}\right].
\end{align*}
\begin{tikzpicture}[remember picture, overlay]
\path[draw=black, thick, <-, dashed] ([yshift = 1.2mm, xshift = 1mm]as) -- ([yshift = 1mm]ae);
\path[draw=black, thick, <-, dashed] ([yshift = 1.2mm, xshift = 1mm]cs) -- ([yshift = 1mm]ce);
\path[draw=black, <-] ([yshift = 1.2mm, xshift = 1mm]bs) -- ([yshift = 1mm]be);
\path[draw=black, <-] ([yshift = 1.2mm, xshift = 1mm]ds) -- ([yshift = 1mm]de);
\end{tikzpicture}
\end{example}

In general, when using the formulation seen in Eq.~\eqref{eq:solution:encoded}, it is going to require exactly $2\;UM$ elements to store an individual, with $U$ unknown contributors and $M$ markers. Compared to the $A_+ U$ elements needed, if used the genotype matrix representation, seen in Eq.~\eqref{eq:solution:genotypematrix}, where $A_+$ is the number of observations in the dataset, i.e.\ $A_+ = \sum_{m = 1}^MA_m$. 

\subsubsection{Fitness of individuals}
We are trying to find unknown genotype combination which maximises the logarithm of the probability of the evidence, i.e.\  
\begin{align*}
\hat{\bs g} = \argmax_{\bs g \; \in \; \mc U} \Big\{ \ell\left(\hat{\bs\theta}_{\bs g} | \bs \beta, \bs\xi, \bs y, \bs g_k, \bs g\right) + \log\left(\p{\bs g|\bs g_k}\right) \Big\},
\end{align*}
where $\bs g_k$ are the known genotypes of the mixture, and $\hat{\bs \theta}_{\bs g}$ is the estimated parameters given $\bs g$ (as well as $\bs g_k$ and $\bs y$). Note: we have inserted the Eq.\ \eqref{eq:app:mixtureloglikelihood}, as the probability of the allelic information. 

We defined the fitness of an individual $\bs p$, corresponding to the unknown genotype(s) $\bs g$, as:
\begin{align}
\begin{split}
&F\left(\bs p \;|\; \hat{\bs \theta}_{\bs g}, \bs y, \bs g_k\right) = \ell\left(\hat{\bs\theta}_{\bs g} \;|\; \bs \beta, \bs\xi, \bs y, \bs g_k, \bs g\right) + \log(\p{\bs g|\bs g_k}),
\end{split}
\end{align}
Because of the definition of the fitness, it follows that any time an individual is changed, we will need to re-estimate the $\bs \theta$ parameters. For notational convenience, we write $F(\bs p)$ instead of $F(\bs p\;|\;\hat{\bs \theta}_{\bs g}, \bs y, \bs g_k)$ from this point forward. 

\subsection{Selection} \label{sec:selection}
Selection is split into two types: parent and survivor selection. The implemented survivor selection is extremely elitist. In order for a child to replace its parent, the fitness of the child has to be larger than the fitness of the parent otherwise the parent survives to the next iteration. 

Parent selection in this implementation works slightly differently than in most EAs, as every individual gets to be a parent and a partner is then selected for that parent. With the partner being selected proportionally to its fitness. Furthermore, to avoid the population fixating a single solution too quickly, we restricted the search of the partner to a neighbourhood around the parent. That is, the probability of an individual $\bs p_j$ being selected as a partner of the parent $\bs p_i$ is defined as:
\begin{align}
\pi^{(s)}(\bs p_j|\bs p_i) = \frac{F(\bs p_j)}{\sum_{k \in \mc N(\bs p_i, L)}F(\bs p_k)},
\end{align}
for all $j \in \mc N(\bs p_i, L)$ and zero otherwise, where $\mc N(\bs p_i, L)$ is the neighbourhood of $\bs p_i$ defined as:
\begin{align}
\mc N(\bs p_i, L) = \Big\{(i + l)\text{ mod } N_{I}\Big\}_{l \in [-L; L] \backslash \{0\}},
\end{align}
i.e.\ a window of size $2L$ with $i$ as its midpoint.

The size of the neighbourhood can be used to control the time to fixation on the sub-population level; the smaller the neighbourhood the longer the time to fixation. 

\subsection{Operators} \label{sec:operators}
\subsubsection{Crossover}
Because of the nature of the implemented parent and survivor selection, we have chosen a crossover operator which creates a single child. The procedure is sketched in Algorithm~\ref{alg:crossover}. Given two individuals, a parent and its partner, the child is created by copying elements one-by-one from either the parent of the partner (the algorithm always starts with parent). After an element has been copied, we switch from parent to partner (or vice versa) with probability of $\pi^{(c)}$, creating a single child of the two individuals. The probability of switching will be inversely proportional with the length of the individual, i.e.\ $\pi^{(c)}\propto 1 / (2\;UM)$. That is, any newly created child will have experienced a single crossover event on average independent of length.

\begin{algorithm}[ht!]
\caption{\label{alg:crossover}Crossover}
\begin{algorithmic}[1]
\renewcommand{\algorithmicrequire}{\textbf{Input:}}
 \renewcommand{\algorithmicensure}{\textbf{Output:}}
 \REQUIRE $\bs p$, $\bs q$, $\pi^{(c)}$
 \ENSURE $\bs c$
 \STATE $s = $ \texttt{false}
  \FOR {$\big(i$ from $0$ to $(2\;UM - 1)\big)$}
  \STATE $u\sim \text{Unif}(0, 1)$.
  \IF {$\left(u < \pi^{(c)}\right)$}
  	\STATE $s = \neg \;s$
  \ENDIF
  \IF {$(s)$} 
  	\STATE $c_i \leftarrow p_i$
  \ELSE
	\STATE $c_i \leftarrow q_i$
  \ENDIF  
  \ENDFOR
 \RETURN $\bs c$ 
\end{algorithmic}
\end{algorithm}

\subsubsection{Mutation}
The mutation operator works on an element-by-element basis choosing to mutate element $i$ with probability $\pi^{(m)}_i$. 

If we choose to mutate an element $c_i$, then the element is changed by drawing a random number $a \in \{1, ..., A_m - 1\}$ and updating the element, as:
\begin{align}\label{eq:mutation}
c_i = (c_i + a)\;\bmod\; A_m
\end{align}
Note that if an element is mutated it always changes its value, as the '0' state is already included as $(1 - \pi_i^{(m)}) > 0$ for all $i$. 

The guided nature of the mutation is controlled through the mutation probabilities. The pseudo-code of the implemented GM is seen in Algorithm~\ref{alg:mutation}. 

Under the probability model given $\bs g$, the observed data $\bs y$ will have an expected value, $\hat{\bs y}$. A standardised residual measures the deviation between these two vectors. Depending on the type of model there exists different types of standardised residual. In particular, the model used is a generalised linear model and we will, therefore, choose the so called deviance residuals, as they are approximately normally distributed. The deviance residuals can be seen in Eq.~\eqref{eq:app:deviance}.

\begin{figure}[ht!]
\begin{minipage}{0.49\textwidth}
\centering
\vspace{22pt}
\includegraphics[width=0.97\columnwidth]{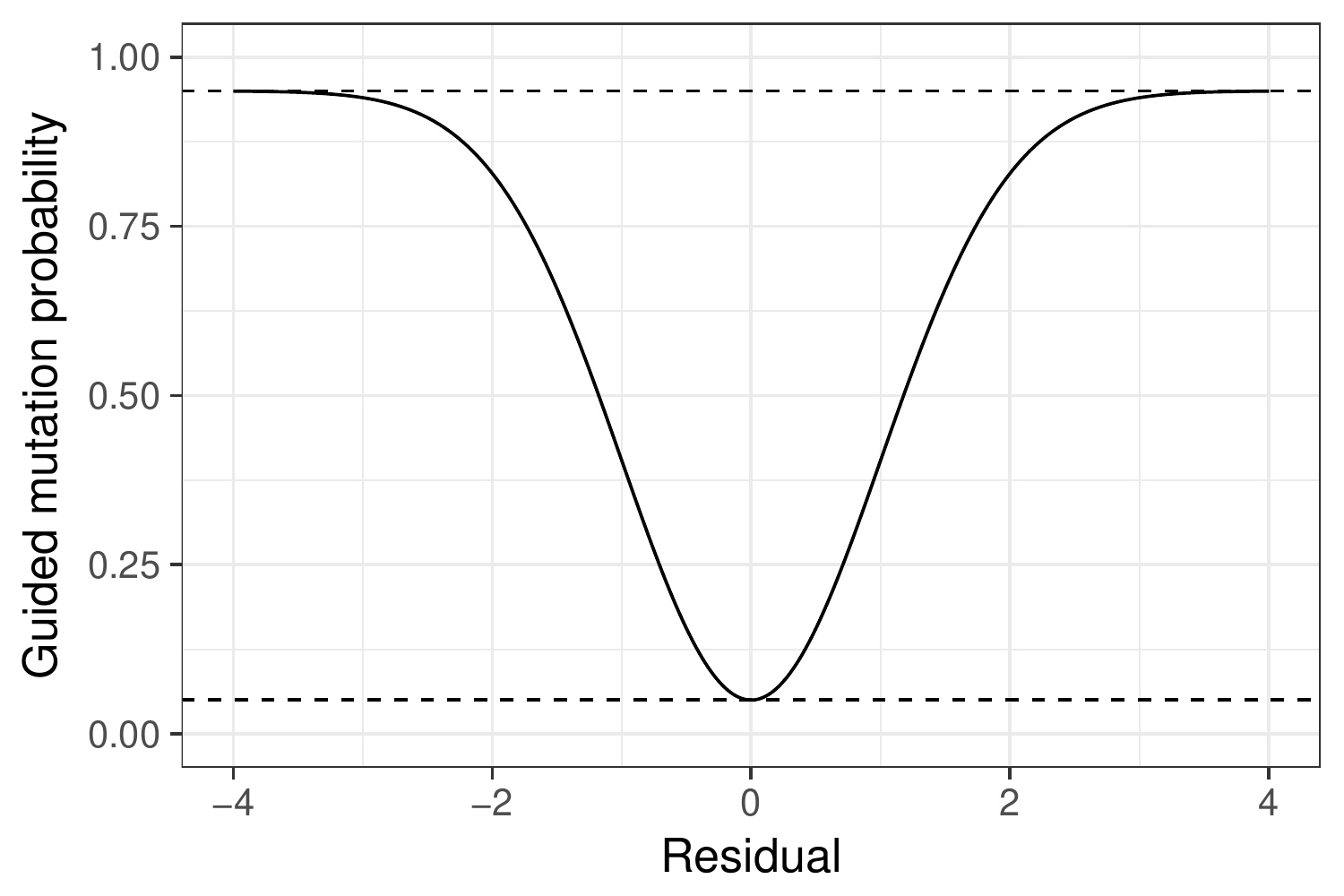}
\end{minipage}
\begin{minipage}{0.49\textwidth}
\centering
\includegraphics[width=\columnwidth]{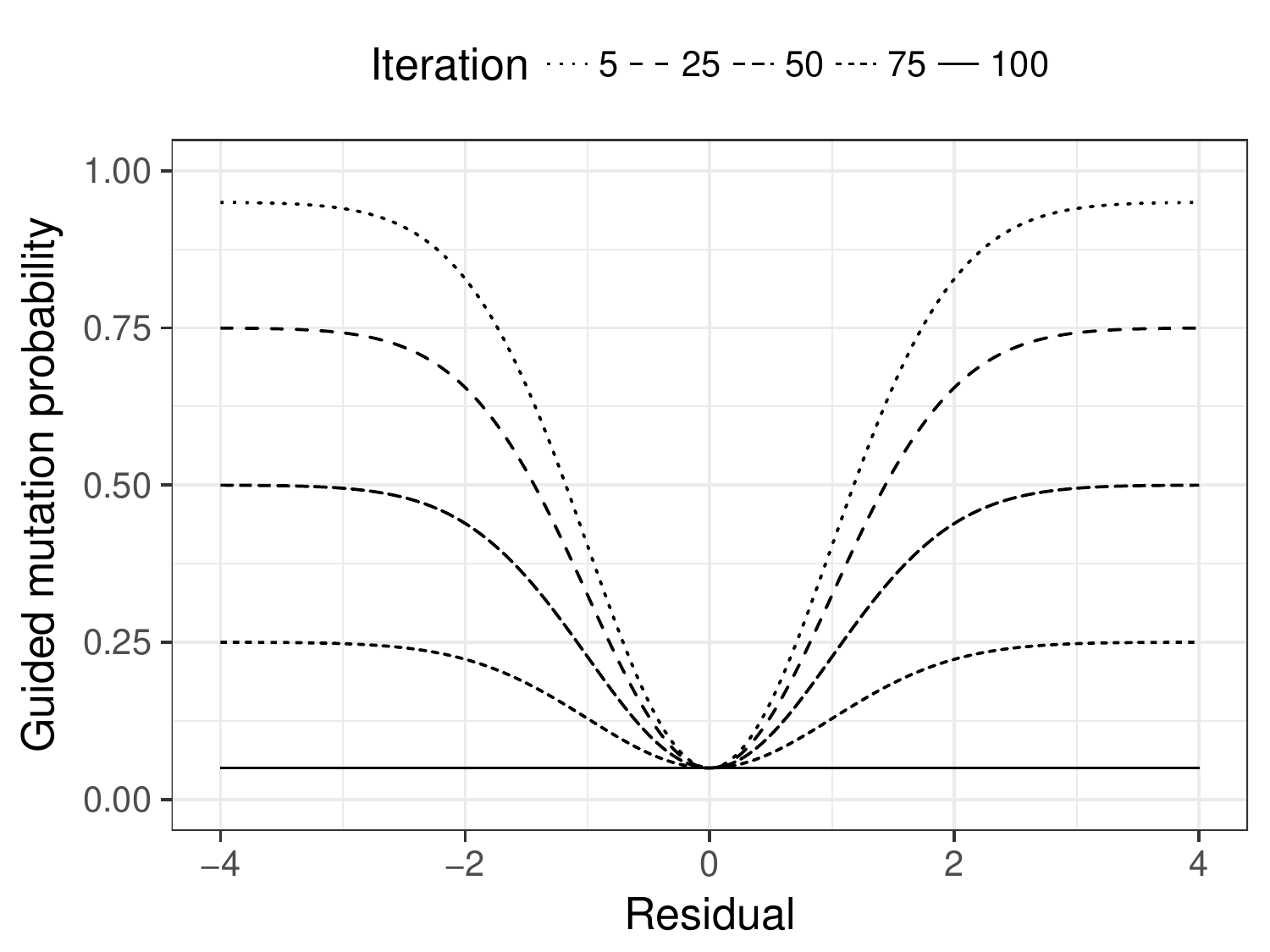}
\end{minipage}
\caption{\label{fig:mutation:prob}The guided mutation probability against the residual. Showing the homogeneous and inhomogeneous on the left and right, respectively. The inhomogeneous case uses an upper bound in time, forcing it to hit the lower bound after 100 iterations.}
\end{figure}

With this in mind, we defined the probability of mutating in the $t$'th iteration of an independent EA, as:
\begin{align}
\begin{split}
&\pi_{i}^{(m)}\left(r_{i}, \pi^{(m)}_{\text{LB}}, \pi^{(m)}_{\text{UB}}\right) \\
&\quad= \begin{array}{ll}
\pi_{\text{UB}}^{(m)} - \left(\pi_{\text{UB}}^{(m)} - \pi_{\text{LB}}^{(m)}\right)\dfrac{f(r_i)}{f(0)}
\end{array}
\end{split}\label{eq:mutation:raw}
\end{align}
where $r_i$ are the deviance residuals (short for $r^D(y_{ma}, \hat \mu_{ma}, \hat \mu_{ma} / \hat \gamma)$), $f$ is the density function of a standard normal distribution, and $\pi^{(m)}_{\text{LB}}$ and $\pi^{(m)}_{\text{UB}}$ are a lower and upper bound on probability of mutation, respectively. The lower and upper bounds were introduced to ensure that the transition matrix was (still) fully connected. That is, they were introduced to ensure that the method still converges towards a global maximum. We have shown a plot of the function on the left-hand side of Fig.~\ref{fig:mutation:prob} with $\pi^{(m)}_{\text{LB}} = 0.05$ and $\pi^{(m)}_{\text{UB}} = 0.95$. 

The probability of mutation described in Eq.~\eqref{eq:mutation:raw} is very useful in the beginning iterations of MEA, but becomes less and less effective as the number of iterations increases. As it will keep mutating the same elements again and again, even if the element is as good as its ever going to get. Therefore, we introduced an iteration dependent upper bound which will tend towards the lower bound as the number of iterations increased. In order to ease notation, we write $\pi^{(m)}_{\text{UB},t}$ and $\pi_{i,t}^{(m)}$ instead of $\pi^{(m)}_{\text{UB}}(t)$ and $\pi_{i}^{(m)}\left(r_{i}, \pi^{(m)}_{\text{LB}}, \pi^{(m)}_{\text{UB},t}\right)$, respectively. The behaviour of the resulting probability of mutation is depicted on the right-hand side of Fig.~\ref{fig:mutation:prob}, using $\pi^{(m)}_{\text{LB}} = 1 - \pi^{(m)}_{\text{UB}, 0} = 0.05$. Note that $\left\{\pi_{\text{UB}, t}^{(m)}\right\}_{t\leq 0}$ can be taken as any non-increasing sequence, with initial value less than or equal to one. Thus, the rate of decay, $x$, can be specified as any strictly positive real number, i.e.\ $x \in \R^+$. Setting the decay rate at $x = 2$ implies that after $N_{\max} / 2$ iterations the guided mutation probability is at $\pi_{\text{LB}}^{(m)}$ for every element of the individual.

\begin{algorithm}[ht!]
\caption{\label{alg:mutation}Guided mutation operator (GM)}
\begin{algorithmic}[1]
\renewcommand{\algorithmicrequire}{\textbf{Input:}}
\renewcommand{\algorithmicensure}{\textbf{Output:}}
\REQUIRE $\bs c$, $\pi_{\text{LB}}^{(m)}$, $\pi_{\text{UB}}^{(m)}(t)$
\ENSURE $\bs c$   
\FOR {$\big(i$ from $0$ to $(2\;UM - 1)\big)$}
	\STATE $m \leftarrow \left\lfloor\dfrac{i}{2\;UM} \right\rfloor$ \\[5pt]
	\STATE $a \leftarrow c_i$
	\STATE $r_i \leftarrow r^D\left(y_{ma}, \hat \mu_{ma}, \dfrac{\hat \mu_{ma}}{\hat \gamma}\right)$ 
	\STATE $\pi_{i,t}^{(m)} \leftarrow \left(1 - \pi_{\text{UB},t}^{(m)}\right) - \left(1 - \pi_{\text{UB}, t}^{(m)} - \pi_{\text{LB}}^{(m)}\right)\dfrac{f(r_i)}{f(0)}$ 
	\STATE $u\sim \text{Unif}(0, 1)$.
	\IF {$\left(u < \pi_{i,t}^{(m)}\right)$}
		\STATE $s \sim \text{Unif}\{1, 2, ..., A_m - 1\}$
		\STATE $c_i \leftarrow (c_i + s) \text{ mod } A_m$
	\ENDIF
\ENDFOR
\RETURN $\bs c$ 
\end{algorithmic}
\end{algorithm}

\subsubsection{Hill-climbing}
We chose to include hill-climbing of the parents in an effort to increase the average fitness of the population even if no of the mutated child reached a higher fitness than the parent. Furthermore, we hill-climb the parent before choosing its partner, thereby, giving the child the best possible outset before mutation.

The guided hill-climbing algorithm, seen in Algorithm~\ref{alg:hillclimbing}, randomly chooses an element of the individual, creates the $(A_m - 1)$ remaining instances of that element and would ideally re-calculate the fitness for every instance. However, this requires re-estimating the parameters, $\bs \theta$, for every instance, which is by far the slowest part of the implementation. 
 
Therefore, we will take a slightly different approach using the raw residuals. The raw residuals are defined as the difference between the observed and expected vectors.

If the raw residuals are negative (positive), then the expected coverage is larger (smaller) than the observed coverage. If they were extremely negative it could imply that the algorithm was expecting a true allele, but is pointing to a what is most likely noise (reversed for extremely positive residuals). In this case, we would like the algorithm to change from pointing at the potential noise, to pointing at a true allele, without having to re-estimate the parameters. 

Our solution was to choose the new instance, $j$, such that $r_{i} = -r_{j}$ (or $r_{i} + r_{j} = 0$). This would ensure that the squared residuals would not change, and, thus, the estimated parameters would not change. In cases where we could not find such an instance, we chose $j$ such that the sum of residuals got as close to zero as possible, i.e.:
\begin{align*}
j = \argmin_{k} |r_{i} + r_{k}|.
\end{align*}

For this instance only, we estimated the parameters and compared the fitness of the new instance to that of the parent. If the fitness of the new instance is larger than that of the parent, then it replaced the parent.

\begin{algorithm}[ht!]
\caption{\label{alg:hillclimbing}Guided hill-climbing}
\begin{algorithmic}[1]
\renewcommand{\algorithmicrequire}{\textbf{Input:}}
\renewcommand{\algorithmicensure}{\textbf{Output:}}
\REQUIRE $\bs p$
\ENSURE $\bs p$   
\FOR {$\big(h$ from $1$ to $N_H\big)$}
	\STATE $i \sim \text{Unif}\{0, 1, ..., 2\;UM - 1\}$	
	\STATE $\mc I$: Empty list of size $A_m - 1$.
	\FOR {$\big(a$ from $1$ to $A_m - 1\big)$}
		\STATE $\bs s \leftarrow \bs p$
		\STATE $s_i \leftarrow (s_i + a) \text{ mod } A_m$
		\STATE $\mc I[a] \leftarrow \bs s$
	\ENDFOR
	\STATE $\bs k \leftarrow \displaystyle\argmin_{\bs s\in \mc I} \Big\{\big|r^D(p_i) + r^D(s_i)\big|\Big\}$
	\IF {$\big(F(\bs p) < F(\bs k)\big)$}
		\STATE $\bs p \leftarrow \bs k$
	\ENDIF
\ENDFOR
\RETURN $\bs p$ 
\end{algorithmic}
\end{algorithm}

\section{Experiments and results}\label{sec:experiments}
\subsection{Data description}
In total 102 samples were quantified. Out of the 102 samples, 30 samples were sequenced using the Illumina MiSeq and ForenSeq Panel B kit \cite{Hussing2018b}, and 72 samples were sequenced using the Ion Torrent S5 and the Precision ID Globalfiler NGS STR panel from Applied Biosystems, ThermoFisher Scientific. The analyses performed below, we have restricted the samples to only the autosomal STR markers.

The 30 samples quantified by the ForenSeq kit, were based on 15 controlled two person mixture experiments sequenced in duplicate, resulting in 30 two-person DNA mixture samples. DNA from two contributors, one male and one female, were used to create the experiments in the following mixture ratios: 1000:1, 100:1, 50:1, 25:1, 12:1, 6:1, 3:1, 1:1, 1:3, 1:6, 1:12, 1:25, 1:50, 1:100, 1:1000. 

The 72 samples quantified by the S5 kit, were based on 8 two person mixtures. The mixtures contained one male and one female donor, and each of the two person mixtures were created in the mixture ratios 10:1, 3:1, and 1:1, with the female donor being the major in all cases. Furthermore, the total amount of input DNA in the mixtures were diluted forming the series 500 pg, 250 pg, and 125 pg. That is, from each two person mixture a total of 24 samples were created.

All experiments were created and sequenced at the Section of Forensic Genetics, Department of Forensic Medicine, Faculty of Health and Medical Sciences, University of Copenhagen, Denmark. 

We used \texttt{STRait Razor} v3.0 \cite{Woerner2017} to (1) identify the STR regions by using unique flanking region sequences \cite{Fordyce2011}, and (2) aggregate the unique sequences. Afterwards, \texttt{STRMPS} \cite{paper4:STRMPS} was used to reduce the number of unique strings, by applying the quality reduction method seen in Appendix \ref{appendix:string:reduction}.

\subsection{Sensitivity study}
We examined the effectiveness of the guided operators and the number of sub-populations used (fixing the total population size) on the deconvolution of the ForenSeq DNA mixtures. In particular, we conducted the following sensitivity experiments:
\begin{itemize}
\item[(1)] Using the RM with the number of hill-climbing iterations at 0 or 2.
\item[(2)] Using the GM with the mutation decay rate equal to $1/2$, $1$, or $2$, and the number of hill-climbing iterations at 0 or 2.
\item[(3)] Using 1, 2, 4, and 8 sub-populations, with the total population size fixed at $200$.
\end{itemize}

In all thirteen cases, we assumed that the minor contributor to the DNA mixture was known (i.e.\ we are looking for the major contributor), measured the time it took to converge, in both iterations and hours, and calculated the average percentage of matching alleles and markers between the optimal unknown major profile and the true major profile. Furthermore, when the mixture proportions approach 1:1, the true major profile will tend to not be the identical to the optimal major profile, as quantifying samples with MPS process, due to its relative immaturity, produces large with-in and between marker variation. Therefore, we also counted the number of times the fitness of the optimal major profile, $F_{\text{optimal}}$, was larger than the fitness of the true major profile, $F_{\text{true}}$. Lastly, this process was repeated 10 times for each of the 30 samples and 13 experiments.

Table~\ref{tab:res:compared} shows the results of the sensitivity studies for Items (1) and (2). The upper and lower part of the table corresponds to using 0 and 2 hill-climbing iterations, respectively. Furthermore, the table displays the average over ten simulations. Starting with the upper half of the table, we see that the speed of the mutation decay does not have much of an effect on the quality of the solutions, as the percentage of correctly assigned alleles was equal across the parameter, and the fitness of the optimal solution was larger than the fitness of the individual comprised of the true profiles of the major and minor contributor. Furthermore, looking at the bottom part of the table, it is clear that adding hill-climbing to the method makes no difference when the mutation is guided, in fact it only makes the MEA slower by a factor of two. However, when the mutation is not guided, we see that MAE only yields adequate results when the mutation is either very skewed, or if hill-climbing is used to augment the algorithm.

\begin{table*}[ht]
\caption{\label{tab:res:compared}Comparing the quality of the solution and the time to convergence, assuming the minor profile was known, when varying the number hill-climbing iterations, and the mutation decay rate. A mutation decay rate of '-' is used to indicate that the mutation was not guided, but completely random.} 
\centering
\resizebox{0.98\hsize}{!}{\begin{tabular}{c|lrrrrrrrrrrrrrrrr}
\toprule
\multicolumn{18}{l}{Hill-climbing iterations $= 0$} \\
\cmidrule(lr){1-4}
\multicolumn{2}{c}{} & \multicolumn{12}{c}{} & \multicolumn{4}{c}{\multirow{2}{*}{\shortstack{Number of times\\ $F_{\text{optimal}} \geq F_{\text{true}}$}}} \\
\multicolumn{2}{c}{} & \multicolumn{4}{c}{Identical alleles (\%)} & \multicolumn{4}{c}{Identical markers (\%)} & \multicolumn{4}{c}{Time (hours)} & \multicolumn{4}{c}{} \\ 
\cmidrule(lr){3-6}\cmidrule(lr){7-10}\cmidrule(lr){11-14}\cmidrule(lr){15-18}
\multicolumn{2}{c}{Mutation decay} & - & $1/2$ & $1$ & $2$ & - & $1/2$ & $1$ & $2$ & - & $1/2$ & $1$ & $2$ & - & $1/2$ & $1$ & $2$ \\
\midrule
\parbox[c]{5pt}{\multirow{8}{*}{\rotatebox[origin=c]{90}{Major:Minor}}} & 1000:1 & 0.85 & 0.99 & 0.99 & 0.99 & 0.79 & 0.98 & 0.98 & 0.98 & 0.42 & 0.35 & 0.34 & 0.34 & 4.00 & 2.00 & 2.00 & 2.00 \\ 
 & 100:1 & 0.69 & 1.00 & 1.00 & 1.00 & 0.61 & 1.00 & 1.00 & 1.00 & 0.34 & 0.27 & 0.27 & 0.28 & 4.00 & 3.80 & 3.90 & 4.00 \\ 
 & 50:1 & 0.66 & 1.00 & 1.00 & 1.00 & 0.59 & 1.00 & 1.00 & 1.00 & 0.28 & 0.23 & 0.22 & 0.22 & 4.00 & 4.00 & 4.00 & 4.00 \\ 
 & 25:1 & 0.63 & 1.00 & 1.00 & 1.00 & 0.56 & 0.99 & 0.99 & 0.99 & 0.42 & 0.29 & 0.30 & 0.29 & 4.00 & 4.00 & 4.00 & 4.00 \\ 
 & 12:1 & 0.67 & 1.00 & 1.00 & 1.00 & 0.60 & 1.00 & 1.00 & 1.00 & 0.28 & 0.25 & 0.24 & 0.24 & 3.00 & 3.00 & 3.00 & 3.00 \\ 
 & 6:1 & 0.64 & 1.00 & 1.00 & 1.00 & 0.57 & 1.00 & 1.00 & 1.00 & 0.27 & 0.26 & 0.26 & 0.25 & 3.00 & 3.00 & 3.00 & 3.00 \\ 
 & 3:1 & 0.56 & 0.99 & 0.99 & 0.99 & 0.52 & 0.97 & 0.97 & 0.97 & 0.26 & 0.25 & 0.25 & 0.25 & 4.00 & 4.00 & 4.00 & 4.00 \\ 
 & 1:1 & 0.55 & 0.96 & 0.96 & 0.96 & 0.48 & 0.91 & 0.91 & 0.91 & 0.35 & 0.32 & 0.33 & 0.33 & 2.00 & 2.00 & 2.00 & 2.00 \\ 
\midrule
 \midrule
 \multicolumn{18}{l}{Hill-climbing iterations $= 2$} \\
\cmidrule(lr){1-4}
\multicolumn{2}{c}{} & \multicolumn{12}{c}{} & \multicolumn{4}{c}{\multirow{2}{*}{\shortstack{Number of times\\ $F_{\text{optimal}} \geq F_{\text{true}}$}}} \\
\multicolumn{2}{c}{} & \multicolumn{4}{c}{Identical alleles (\%)} & \multicolumn{4}{c}{Identical markers (\%)} & \multicolumn{4}{c}{Time (hours)} & \multicolumn{4}{c}{} \\ 
\cmidrule(lr){3-6}\cmidrule(lr){7-10}\cmidrule(lr){11-14}\cmidrule(lr){15-18}
\multicolumn{2}{c}{Mutation decay} & - & $1/2$ & $1$ & $2$ & - & $1/2$ & $1$ & $2$ & NA & $1/2$ & $1$ & $2$ & - & $1/2$ & $1$ & $2$ \\
\midrule
\parbox[c]{5pt}{\multirow{8}{*}{\rotatebox[origin=c]{90}{Major:Minor}}} & 1000:1 & 0.99 & 0.99 & 0.99 & 0.99 & 0.98 & 0.98 & 0.98 & 0.98 & 1.30 & 0.51 & 0.50 & 0.49 & 4.00 & 2.00 & 2.00 & 2.00 \\ 
  & 100:1 & 1.00 & 1.00 & 1.00 & 1.00 & 1.00 & 1.00 & 1.00 & 1.00 & 1.55 & 0.54 & 0.55 & 0.53 & 8.00 & 4.00 & 4.00 & 3.90 \\ 
  & 50:1 & 1.00 & 1.00 & 1.00 & 1.00 & 1.00 & 1.00 & 1.00 & 1.00 & 1.67 & 0.46 & 0.44 & 0.46 & 7.80 & 4.00 & 4.00 & 4.00 \\ 
  & 25:1 & 1.00 & 1.00 & 1.00 & 1.00 & 0.99 & 0.99 & 0.99 & 0.99 & 1.59 & 0.57 & 0.58 & 0.56 & 8.00 & 4.00 & 4.00 & 4.00 \\ 
  & 12:1 & 1.00 & 1.00 & 1.00 & 1.00 & 1.00 & 1.00 & 1.00 & 1.00 & 1.76 & 0.47 & 0.46 & 0.45 & 6.00 & 3.00 & 3.00 & 3.00 \\ 
  & 6:1 & 1.00 & 1.00 & 1.00 & 1.00 & 1.00 & 1.00 & 1.00 & 1.00 & 1.33 & 0.49 & 0.50 & 0.49 & 6.00 & 3.00 & 3.00 & 3.00 \\ 
  & 3:1 & 0.99 & 0.99 & 0.99 & 0.99 & 0.97 & 0.97 & 0.97 & 0.97 & 1.43 & 0.55 & 0.53 & 0.57 & 8.00 & 4.00 & 4.00 & 4.00 \\ 
  & 1:1 & 0.96 & 0.96 & 0.96 & 0.96 & 0.91 & 0.91 & 0.91 & 0.91 & 0.99 & 0.68 & 0.75 & 0.72 & 4.00 & 2.00 & 2.00 & 2.00 \\
\bottomrule
\end{tabular}}
\end{table*}

The results of the sensitivity study described in Item (3) are shown in Table~\ref{tab:res:populationCompared}. We see that the number of sub-populations does not have much of an influence on the quality of the solutions, though only having a single population seems to cause a little trouble for two of the mixture proportions. The largest effect of the using multiple sub-populations is seen in the time to completion, which drastically decreases as we increased the number sub-populations. This decrease is a combination of two factors: (1) each sub-population can be run in parallel, and (2) the sub-populations a randomly initialised so they more effectively search the state-space. Furthermore, we see that the computational increases slightly when going from a single population to two sub-populations, this is likely due to the added overhead of running multiple populations (e.g.\ set-up of parallel processes, migration between sub-populations, and so on).

\begin{table*}[ht!]
\caption{\label{tab:res:populationCompared}Comparing the quality of the solution and the time to convergence, assuming the minor profile was known, when varying the number of sub-populations while keeping the total population size fixed at $200$. Note: the number of iterations are not shown for the single population cases, as it is not directly comparable to the multiple population cases.}
\centering
\resizebox{0.98\hsize}{!}{\begin{tabular}{c|lrrrrrrrrrrrr}
  \toprule
 \multicolumn{2}{c}{} & \multicolumn{8}{c}{} & \multicolumn{4}{c}{\multirow{2}{*}{\shortstack{Number of times\\ $F_{\text{optimal}} \geq F_{\text{true}}$}}} \\
\multicolumn{2}{c}{} & \multicolumn{4}{c}{Time (iterations)} & \multicolumn{4}{c}{Time (hours)} & \multicolumn{4}{c}{} \\ 
\cmidrule(lr){3-6}\cmidrule(lr){7-10}\cmidrule(lr){11-14} 
\multicolumn{2}{c}{\# Sub-populations} & 1 & 2 & 4 & 8 & 1 & 2 & 4 & 8 & 1 & 2 & 4 & 8 \\  
  \midrule
\parbox[c]{5pt}{\multirow{8}{*}{\rotatebox[origin=c]{90}{Major:Minor}}} & 1000:1 & - & 47.40 & 45.40 & 50.80 & 1.30 & 1.71 & 0.89 & 0.46 & 2.00 & 2.00 & 2.00 & 2.00 \\ 
  & 100:1 & - & 42.10 & 44.20 & 48.70 & 1.55 & 1.59 & 0.89 & 0.52 & 3.30 & 4.00 & 4.00 & 3.90 \\ 
  & 50:1 & - & 41.80 & 45.80 & 49.30 & 1.54 & 1.61 & 0.81 & 0.46 & 4.00 & 4.00 & 4.00 & 4.00 \\ 
  & 25:1 & - & 41.30 & 45.90 & 49.50 & 1.81 & 1.81 & 1.02 & 0.58 & 3.10 & 4.00 & 4.00 & 4.00 \\ 
  & 12:1 & - & 39.90 & 44.80 & 48.50 & 1.46 & 1.51 & 0.84 & 0.47 & 3.00 & 3.00 & 3.00 & 3.00 \\ 
  & 6:1 & - & 45.00 & 45.60 & 52.30 & 1.47 & 1.67 & 0.88 & 0.52 & 3.00 & 3.00 & 3.00 & 3.00 \\ 
  & 3:1 & - & 43.20 & 45.70 & 49.60 & 1.38 & 2.01 & 0.89 & 0.50 & 4.00 & 4.00 & 4.00 & 4.00 \\ 
  & 1:1 & - & 44.50 & 48.00 & 52.90 & 1.99 & 2.06 & 1.09 & 0.62 & 2.00 & 2.00 & 2.00 & 2.00 \\  
   \bottomrule
\end{tabular}}
\end{table*}

Based on the sensitivity study, we will from this point forward use the following settings: 16 sub-populations, 125 individuals per sub-populations, 10 inner iterations, 250 outer iterations, mutation decay rate of 1, and no hill-climbing. Note that the number of sub-populations and number of individuals per sub-populations are slightly higher than in the sensitivity study, sacrificing speed for slightly better results.

\subsection{Deconvolution}
For each of the 102 samples, we conducted three deconvolution experiments: (1) assuming the minor profile was known, (2) assuming the major profile was known, and (3) assuming both profiles were unknown. In all three experiments, we counted the number of alleles the optimal profile had in common with the true profile. The correct allele percentage was then compared across kits, amount of total input DNA, and the intended mixture proportions.

\begin{figure}[ht!]
\centering
\includegraphics[page = 2, width = 0.7\columnwidth]{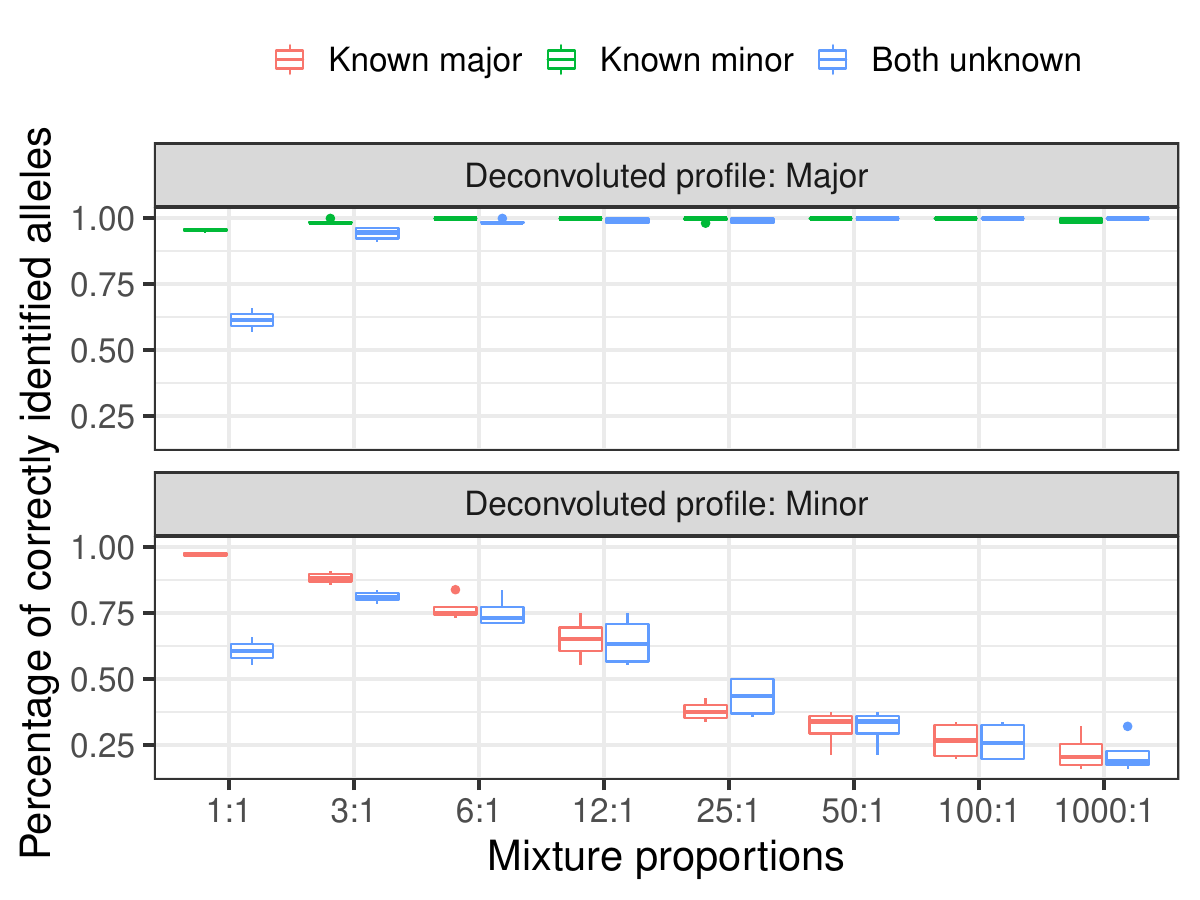}
\caption{\label{fig:deconvolution:forenseq}Boxplots of the percentage of correctly identified alleles against the mixture proportion of the true profile. The top panel shows the results for the major profile, while the bottom panel shows the minor profile. The red, green, and blue colours corresponds to the hypotheses of the major profile is known, the minor profile is known, and both the profiles are unknown, respectively.}
\end{figure}

The results of the deconvolution experiments for the Illumina ForenSeq panel are shown in Fig.~\ref{fig:deconvolution:forenseq}. It shows boxplots of the percentage of correctly identified alleles when the optimal profile deconvoluted by the algorithm is compared to the true profiles of the sample. The top panel of the figure shows the results for the major profile (i.e.\ when either the minor profile was assumed known, or when both profiles were assumed unknown). We see that when the minor profile was assumed known the major profile was identified at a very high accuracy (larger than 97\% in all cases), while when both the profiles were unknown the percentage of correctly identified alleles decreases slightly as the mixture proportions tend towards a 1:1 mixture (or at least as it gets less skewed than a 3:1 mixture). The bottom panel of the figure shows the results for the minor profile (i.e.\ when the major profile was known or both were unknown). We see that as the mixture proportions become more skewed, i.e.\ tending from 1:1 to a 1000:1 mixture, it becomes more and more difficult to identify the minor profile. This is natural, as the mixture proportion gets more skewed, but the total amount of input DNA is fixed, the alleles of the minor profile is drowned in the noise created by the MPS process. However, it is worth noting that at a 100:1 mixture 25\% of the correct alleles are still identified. When both profiles were assumed unknown the boxplots are almost identical when the mixture is more skewed than 3:1, but looking at the 1:1 and even 3:1 mixtures, we see a drop similar to what we saw for the major profiles. This is an effect of the independence assumption between markers, i.e.\ for a 1:1 mixture the markers become interchangeable between the contributors, hence, we see a drop in performance (and should expect to, when performing a direct comparison).

\begin{figure}[ht!]
\centering
\includegraphics[page = 7, width = 0.8\columnwidth]{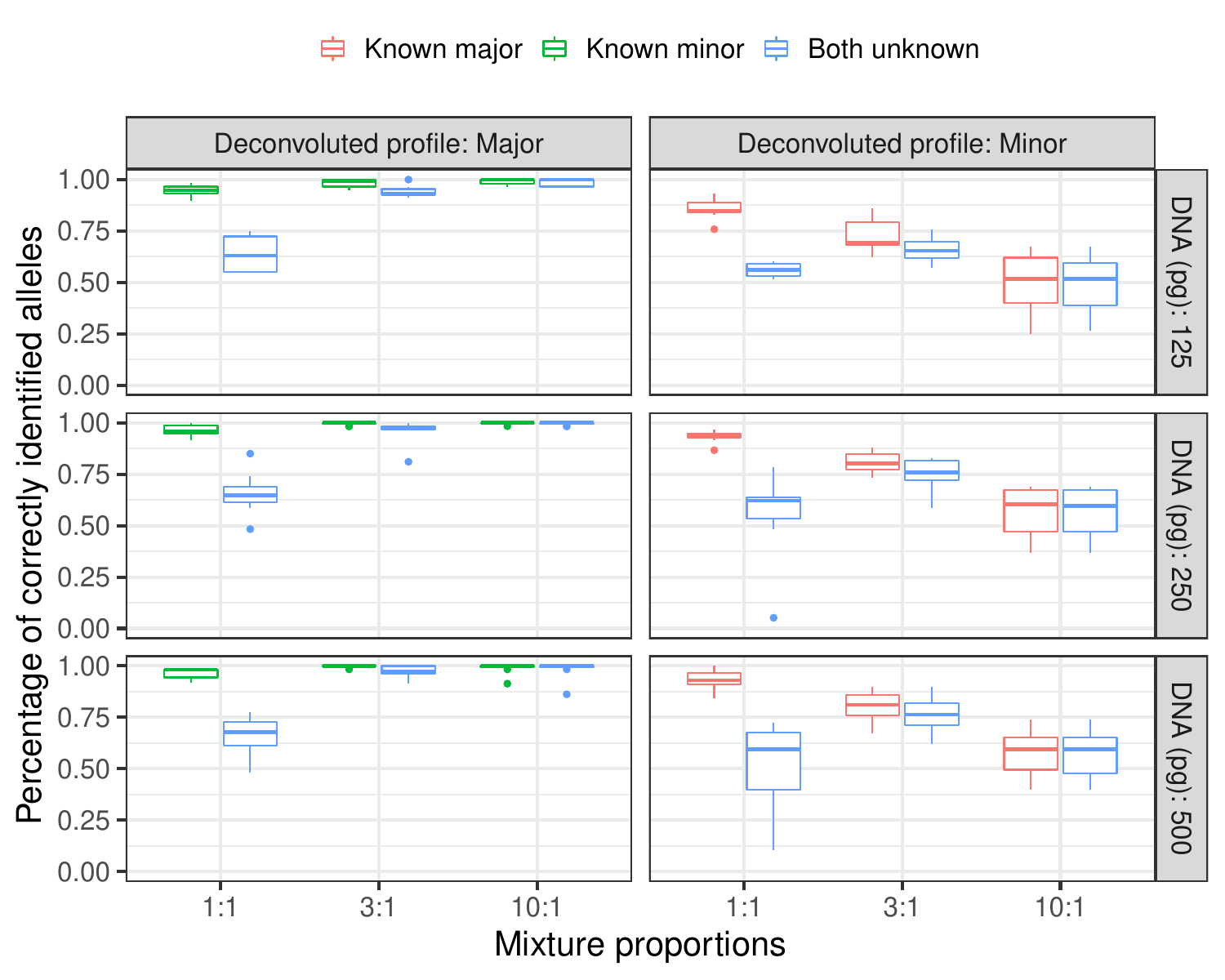}
\caption{\label{fig:deconvolution:s5}The percentage of correctly identified alleles against the mixture proportion of the true profile. The columns correspond to the amount of input DNA (125 pg, 250 pg, and 500 pg), and the rows show the results for deconvoluted major and minor profiles, respectively. The red, green, and blue colours corresponds to the hypotheses of the major profile is known, the minor profile is known, and both the profiles are unknown, respectively.}
\end{figure}

The deconvolution results for the Precision ID Globalfiler panel are shown in Fig.~\ref{fig:deconvolution:s5}. The figure shows a the result to the the percentage of correctly identified alleles when the optimal profile deconvoluted by the algorithm is compared to the true profiles of the sample. This is further stratified by the amount of total input DNA in the sample. Generally the results were very similar to that of the ForenSeq kit, seen in Fig.~\ref{fig:deconvolution:forenseq}. The additional information gained from the figure is the effect of the amount of input DNA on the precision of the deconvolution. The precision when identifying the major profile does not change as the amount of input DNA increases. However, looking at the column showing the precision of the deconvolution of the minor profile, we see that as the amount of input DNA increase the deconvolution precision increases, when the major profile is known. If both are unknown there does not seem to be any difference in the precision as the amount of input DNA increases.

\section{Discussion}\label{sec:conc}
The sensitivity studies show a very stable method across the number of populations, the mutation decay rate, and the number of hill-climbing iterations. The study shows that guiding the mutation in the early stages of the algorithm was beneficial for the time to convergence of the MEA, in some cases decreasing the computational time by more than a factor of two. Furthermore, we saw that while using hill-climbing did little else than increase the computational time, when using the GM operator, it was shown to be useful when employing a RM operator. The added computational time is a simple consequence of having to re-estimate the parameters of the new instance proposed during each hill-climbing iteration. Thus, when using two hill-climbing iterations, we need to re-estimate the parameters an additional two times, per individual, per iteration, and as parameter estimation is by far the slowest part of the algorithm it greatly increases the computational time.

In general the results of the deconvolution were as expected; when minor profile is known identifying the major profile was simple for any mixture proportion. While the precision of the identified minor profile, when the major profile was known, increased as the mixture proportion tended towards a 1:1 mixture. When both profiles were unknown, the precision of the deconvolution of the major profile increased as the mixture got more skewed. While the precision of the deconvolution of the minor profile increased slightly when the mixture proportions increase from 1:1 to 3:1, it then decreases as the skewness of the mixture increases. Furthermore, we saw that as the amount of input DNA decreased the precision of the deconvolution of the minor profile decreased, when the major profile was known. With that said, the decrease in precision is very slight, as the amount of input DNA is still large enough to not induce much dropout in the minor profile, even for the very skewed 10:1 mixture. In fact, the precision of the deconvoluted minor profile in samples with 10:1 mixture proportion, does not seem to change as the amount of DNA decreases.

\section*{Acknowledgment}
The authors would like to thank associate professor Leif K. J{\o}rgensen for his help in deriving the minimum number of iterations needed for an individual to spread to every sub-population.

\bibliographystyle{unsrtnat}
\bibliography{Literature.bib}

\begin{appendices}
\section{The Poisson-gamma distribution}\label{app:poissongamma}
If $Y$ follows a Poisson-gamma distribution (also known as the mean parameterised negative binomial distribution or the negative binomial regression model) with parameters $\mu$ and $\eta$, we write $Y \sim \text{PG}(\mu, \eta)$. Furthermore, the probability of $\p{Y = y|\mu, \eta}$ is:
\begin{align*}
p(y|\mu, \eta) = \frac{\Gamma(y + \eta)}{\Gamma(y + 1)\Gamma(\eta)}\frac{\mu^y\eta^\eta}{(\mu + \eta)^{y + \eta}}.
\end{align*}

The expectation and variance of $Y$ are:
\begin{align}
\E{Y} &= \mu\;\;\text{and} \nonumber \\ 
\Var{Y} &= \mu \left(1 + \frac{\mu}{\eta}\right),\label{eq:app:pg:pg2var}
\end{align}
respectively.

Assuming $y_n \sim \text{PG}(\mu_n, \eta_n)$ for $n = 1,..., N$ and that $y_n$ and $y_m$ are independent, then the log-likelihood is given as: 
\begin{align}
\begin{split}
&\ell(\bs \mu, \bs \eta|\bs y) = \\
&\;\;\sum_n \log\left(\Gamma(y_n + \eta_n)\right) - \log\left(\Gamma(y_n + 1)\right) \\ 
&\;\;- \log\left(\Gamma(\eta_n)\right) + y_n \log(\mu_n) + \eta\log(\eta_n) \\
&\;\;- (y_n + \eta_n)\log(\mu_n + \eta_n).
\end{split}\nonumber
\end{align}

Furthermore, assuming $\hat{\bs \mu}$ and $\hat{\eta}$ are the MLEs of $\bs\mu$ and $\eta$, respectively, we can write the deviance residuals of observation $n$ as:
\begin{align}
\begin{split}
&r^D(y_n, \hat{\mu}_n, \hat{\eta}_n) = \\ 
&\quad\text{sign}(y_n - \hat{\mu}_n)\left(2\left[(y_n + \hat \eta_n)\log\left(\frac{\hat \mu_n + \hat \eta_n}{y_n + \hat \eta_n}\right)\right. \right. \\ 
&\qquad\qquad\qquad\qquad\left. \left. +\; y_n\log\left(\frac{y_n}{\hat \mu_n}\right)\right]\right)^{1 / 2}.
\end{split}\label{eq:app:deviance}
\end{align}

This variant of the Poisson-gamma distribution is also called the PG2 model, because the variance is dominated by the expectation to the power of 2. Instead of the using the PG2 model, we will use the PG1 variant. In order to derive the PG1 model, assume that the overdispersion of the PG2 model, $\eta_n$, is given as:
\begin{align*}
\eta_{n} = \frac{\mu_{n}}{\gamma}.
\end{align*}

If $\eta_{n}$ is inserted in Eq.~\eqref{eq:app:pg:pg2var}, we get:
\begin{align*}
\Var{Y_n} = \mu_{n} (1  + \gamma),
\end{align*}
i.e.\ the variance is now proportional to $\mu$ (giving it the name PG1).

\section{Reducing the number of base-calling errors}\label{appendix:string:reduction}
Our aim was to avoid thresholding by modelling the noise separately from the allele signals and the systematic errors (primarily stutters). Identifying the STR regions of e.g.\ a ForenSeq-kits \texttt{fastq} file with approximately $300{,}000$ to $400{,}000$ reads by searching for flanking regions (by using STRaitRazor version 3 \cite{Woerner2017}) results in approximately $4{,}000$ to $6{,}000$ unique strings. Restricting ourselves to the autosomal markers and assuming a single contributor, then approximately one hundred of these strings would be either alleles or systematic errors (leaving approximately $3{,}900$ to $5{,}900$ unique strings contributed to general noise). An important question is: can the number of remaining strings (i.e.\ the noise) be reduced? That is, could we determine from which '\emph{true}' strings the noise originated? And could we use this information to increase the coverage of the '\emph{true}' strings using the coverage of the strings found among the sequencing errors?

The reduction in the number of unique strings will be based on the base qualities provided in the \texttt{fastq} files by analysing the probability of bases having been called erroneously. In the remainder of this appendix, we will for simplicity restrict the discussion to a single marker and strings of equal length. As the aim is to distinguish isoalleles from base calling errors.

\subsection{The quality of a base}
The quality of a sequenced/called base is defined as a transformation of the probability of the base having been called erroneously. Today, the most common definition of the quality is the Phred score. Given the probability that base $n$ was called erroneously, $p_n$, the Phred score is defined as:
\begin{align}
q_n = -10\log_{10}(p_n).
\end{align} 

Thus, given quality, $q_n$, the probability, $p_n$, is easily recovered:
\begin{align}
p_n = 10^{-q_n/10}.
\end{align}

Given a string $\bs s$ (seen as a vector of characters i.e.\ bases) of length $N$ and the corresponding vector of probabilities $\bs p$, we want to find the probability that every base of $\bs s$ was called correctly. If the bases could be considered independent, this probability would simply be:
\begin{align}\label{eq:probcorrect}
\p{\bs s \text{ is correct}} = \prod_{n = 1}^N (1 - p_n).
\end{align}

However, if a base is called erroneously, it may not necessarily be reflected in the quality of that base, but it may affect the quality of the surrounding bases due to a sliding window. Thus, the base qualities are not independent. Therefore, we will instead of using $p_n$ directly compensate for the quality of the surrounding bases by using $\pi_n$, which is defined as:
\[
\pi_n = \displaystyle\max\limits_{j \in \bar{\delta}(n, l)}\left\{\frac{p_j}{|j - i| + 1}\right\},
\]
where $\bar{\delta}(n, l)$ is a neighbourhood of $n$, including $n$ itself, of size at most $2l$ bases. That is, we weight the probabilities in a neighbourhood around base $n$ by its distance to $n$ and take the largest weighted probability as $\pi_n$. It follows that the probabilities calculated below cannot be interpreted as actual probabilities. That is, we have taken an entirely heuristic approach.

Given two strings $\bs s_i$ and $\bs s_j$, we want to calculate the probability of $\bs s_i$ actually being the string $\bs s_j$ with some number of miscalled bases denoted $\bs s_i \equiv \bs s_j$:
\begin{align}
\begin{split}
&\p{\bs s_i \equiv \bs s_j \text{ and } \bs s_j\text{ is correct}} = \\ 
&\qquad \p{\bs s_i \equiv \bs s_j | \bs s_j\text{ is correct}} \p{\bs s_j\text{ is correct}}.
\end{split} \label{eq:probvariant}
\end{align}

Assuming we know the probability of error of each base of the string $\bs s_i$, denoted $\bs p_i$, then the conditional probability in Eq.~\eqref{eq:probvariant} can be written as:
\begin{align} \label{eq:probvariantconditional}
\p{\bs s_i \equiv \bs s_j | \bs p_i, \bs s_j} = \prod_{n = 1}^N (1 - \pi_{in})^{1 - I_n}(\pi_{in})^{I_n},
\end{align}
where $I_n = \mathbb{I}\left[s_{in} \neq s_{jn}\right]$, i.e.\ a function indicating whether $s_{in}$ is equal to the '\emph{truth}' $s_{jn}$. 

The probability that the string '\emph{is correct}' could be based entirely on the quality, in which case it would be given by Eq.~\eqref{eq:probcorrect}. However, defining it in this way would entirely ignore the information found in the coverage. Therefore, we have defined the probability of a string being correct as:
\begin{align}
\begin{split}
&\p{\bs s_j\text{ is correct}} = \\ 
&\qquad w_j\p{\bs s_j\text{ does not contain base errors}},
\end{split} \label{eq:probcorrectfull}
\end{align}
where $\p{\bs s_j\text{ does not contain base errors}}$ is given directly by Eq.~\eqref{eq:probcorrect} and the weight, $w_j$, of $\bs s_j$ is given by:
\begin{align}
w_j = \frac{y_{j}}{\displaystyle\sum\limits_{k = 1}^K y_k},
\end{align}
where $y_k$ is the coverage of string $\bs s_k$ and $K$ is the number of strings of length $N$. This further emphasises that our approach is heuristic and not analytic.

\subsection{Reduction approach}
At a given STR marker, assume we have observed a set of strings of length $N$, $\bs S = (\bs s_1, \bs s_2, ..., \bs s_K)$ and the coverage of each string given by $\bs y = (y_1, y_2, ..., y_K)$. A matrix of joint probabilities is constructed as given by Eq.~\eqref{eq:probvariant} denoted $V \in \mathbb{R}^{K\times K}$. Given the matrix $V$, we find the index of largest probability of every column, i.e.\ 
\[
k(i) = \argmax_j \{v_{ji}\}.
\]

As the largest probability of the $i$'th column is $k(i)$, we say that the string $\bs s_{i}$ is actually the string $\bs s_{k(i)}$, but called with errors, and we remove $\bs s_{i}$ from further consideration. In cases where the coverage is low, it can be beneficial to retain the information of the coverage of the string $\bs s_{i}$ and add the coverage of $\bs s_i$ to the coverage of $\bs s_{k(i)}$. The probability matrix $V$ can be constructed by calculating every pairwise combination of Eq.~\eqref{eq:probvariant} for the strings in $\bs S$. 

However, we are generally not interested in finding new variants when analysing mixtures. Therefore, we propose the following scheme using a database of '\emph{true}' (or '\emph{trusted}') STR variants.

Assuming we have database of '\emph{true}' STR variants (of the specified marker and length) denoted $\mc T = (\bs t_1, \bs t_2, ..., \bs t_M)$. 
At this point, we can take the problem in two directions dependent on whether we want to allow for the survival of any variants. If we do not want this, we just want to calculate the probability of every string in the set $\bs S$ being a variant of a string in the set $\bs S \cap \mc T$. If we want to allow for the survival of variant strings, we still need the first part, but in addition, we also want the probability of the strings not in $\bs S \cap \mc T$ (i.e.\ $\bs S \backslash \mc T$) being called correctly. Note: if $\bs S \cap \mc T = \emptyset$, then we would have to calculate every possible pairwise combination.

By the definition of $\mc T$, it follows that $\p{\bs s_i\text{ is correct}} = 1$ for all $\bs s_i \in \bs S \cap \mc T$. Thus, calculating the joint probability in Eq.~\eqref{eq:probvariant} is reduced to calculating the conditional probability in Eq.~\eqref{eq:probvariantconditional}. That is, we calculate $\p{\bs s_i \equiv \bs s_j | \bs p_i, \bs s_j}$ for every $\bs s_i \in \bs S$ and $\bs s_j \in \bs S \cap \mc T$, creating a matrix, $V^{(1)}$, of size $I \times N$, where $I$ is the number of elements in $\bs S \cap \mc T$ (i.e.\ $I = |\bs S \cap \mc T|$). Assuming, without loss of generality, that the matrix (and the set $\bs S$) is ordered such that the first $I$ columns (elements) corresponds to the strings in $\bs S \cap \mc T$, then the matrix will have the following structure:
\[
\resizebox{0.98\hsize}{!}{$\displaystyle
V^{(1)} = \begin{bmatrix}
1 & 0 & \cdots & 0 & \p{\bs s_{I + 1} \equiv \bs s_1 | \bs p_{I + 1}, \bs s_1} & \p{\bs s_{I + 2} \equiv \bs s_1 | \bs p_{I + 2}, \bs s_1} & \cdots & \p{\bs s_{N - I} \equiv \bs s_1 | \bs p_{N - I}, \bs s_1} \\
0 & 1 & \cdots & 0 & \p{\bs s_{I + 1} \equiv \bs s_2 | \bs p_{I + 1}, \bs s_2} & \p{\bs s_{I + 2} \equiv \bs s_2 | \bs p_{I + 2}, \bs s_2} & \cdots & \p{\bs s_{N - I} \equiv \bs s_2 | \bs p_{N - I}, \bs s_2} \\
\vdots & \vdots & \ddots & \vdots & \vdots & \vdots & \ddots & \vdots \\
0 & 0 & \cdots & 1  & \p{\bs s_{I + 1} \equiv \bs s_{I} | \bs p_{I + 1}, \bs s_{I}} & \p{\bs s_{I + 2} \equiv \bs s_{I} | \bs p_{I + 2}, \bs s_{I}} & \cdots & \p{\bs s_{N - I} \equiv \bs s_{I} | \bs p_{N - I}, \bs s_{I}}
\end{bmatrix}.$}
\]

If we do not want to allow for any variants, then we set $V = V^{(1)}$ and find the '\emph{true}' string for every string in $\bs S$ as described above. 

If we want to allow the variant strings (i.e.\ strings in $\bs S \backslash \mc T$) a chance to survive, we need the probabilities of these strings having been called correctly. Note we are not interested in the probability of the strings in $\bs S \cap \mc T$ are called correctly, as they are assumed correct given $\mc T$. Furthermore, we were not interested in the probability of a variant string was in fact just another variant string having miscalled bases. 

Thus, for each string in $\bs S \backslash \mc T$, we calculate the probability of the string was called correctly, using Eq.~\eqref{eq:probcorrectfull}. Assuming, as before, that the first $I$ columns corresponds to the strings in $\bs S \cap \mc T$, we have defined a matrix $V^{(2)}$ of size $(K - I) \times K$, taking the form $V^{(2)} = [O \; J]$, where $O$ is a $I\times I$ matrix of zeroes and $J$ is the diagonal matrix $\{\text{diag}(\p{\bs s_{j}\text{ is correct}})\;|\; j\in \bs S \backslash \mc T\}$. The matrix $V$ is then constructed as:
\[
V = \begin{bmatrix}
V^{(1)} \\
V^{(2)}
\end{bmatrix},
\]
and the index of the largest probability is found as discussed above. 

Using this approach, we can reduce the number of unique strings from $4{,}000$-$6{,}000$ to less than $500$. At the same time, we keep the coverage of the sample the same by adding the coverage of the strings in $\bs S \backslash \mc T$ having been assigned a '\emph{true}' string in $\bs S \cap \mc T$ to the '\emph{true}' string. 

\end{appendices}
\end{document}